\documentclass[aps,prb,twocolumn,groupedaddress,showpacs]{revtex4-1}
\usepackage{graphicx}
\usepackage{amsmath}
\usepackage{amssymb}
\usepackage{bm}
\usepackage{dcolumn}
\usepackage{subfigure}
\usepackage{psfrag}
\usepackage{float}
\usepackage{subeqnarray}
\usepackage{url}
\usepackage{units}

\begin{document}

\title{The Tomasch effect in nanoscale superconductors}

\author{L.-F. Zhang}\email{lingfeng.zhang@uantwerpen.be}
\author{L. Covaci}\email{lucian@covaci.org}
\author{F. M. Peeters}\email{francois.peeters@uantwerpen.be}
\affiliation{Departement Fysica, Universiteit Antwerpen,
Groenenborgerlaan 171, B-2020 Antwerpen, Belgium}

\date{\today}

\begin{abstract}
The Tomasch effect (TE) is due to quasiparticle interference (QPI) as induced by a
nonuniform superconducting order parameter, which results in oscillations in the density of
states (DOS) at energies above the superconducting gap.  Quantum confinement in nanoscale
superconductors leads to an inhomogenerous distribution of the Cooper-pair condensate,
which, as we found, triggers the manifestation of a new TE.  We investigate the electronic structure of nanoscale superconductors by solving
the Bogoliubov-de Gennes (BdG) equations self-consistently and describe the TE determined by two types of
processes, involving two- or three- subband QPIs. Both types of QPIs result in
additional BCS-like Bogoliubov-quasiparticles and BCS-like energy gaps leading to oscillations in
the DOS and modulated wave patterns in the local density of states.  These effects are
strongly related to the symmetries of the system. A reduced $4 \times 4$ inter-subband BdG
Hamiltonian is established in order to describe analytically the TE of two-subband QPIs. Our study is
relevant to nanoscale superconductors, either nanowires or thin films, Bose-Einsten condensates and
confined systems such as two-dimensional electron gas interface superconductivity.
\end{abstract}



\maketitle
\section{Introduction}
Electronic structure has always been one of the most important topics in understanding transport
properties in condensed matter.  A variety of remarkable phenomena, from traditional conducting,
semiconducting or insulating behavior to contemporary quantum Hall effect\cite{QHE} and topological
insulator behavior\cite{TI}, are induced by a diversity of electronic structures.

For conventional superconductors, the well-known BCS theory\cite{BCS} and its generalization, the Bogoliubov-de Gennes (BdG) equations\cite{DeGennes}, are the milestones needed to reveal the electronic structure theoretically.  An energy gap created around the Fermi level, $E_F$, in homogeneous superconductivity and low-lying bound states existing in the core of vortex states\cite{vortex1, vortex2, vortex3, vortex4, vortex5, vortex6, vortex7} were successfully predicted and coincide with experimental results from tunneling conductance and scanning tunneling microscopy (STM)\cite{vortexp1, vortexp2, vortexp3, vortexp4}.

Nanoscale superconductors have also received considerable attention in the last decades due to
developments in nanotechnology and their unique properties such as shell effect,\cite{bose_shell}
quantum-size effect\cite{QSEXP1, QSEXP2, QSE1, QSE2,QSE3} and quantum-size cascades under magnetic
field\cite{QSCascade}.  All these effects are induced by changes in the electronic structure
resulting from quantum confinement, such that energy levels are discretized in nanoparticles and
single-electron subbands appears in nanowires.  In addition, such electronic structure results in a
spatially inhomogeneous superconducting order parameter\cite{QSE2}, which further induces other
effects such as superconducting multi-gap structures,\cite{MultiGap, NAV} new
Andreev-states,\cite{NAV} unconventional vortex states,\cite{LFZ} and a position-dependent impurity
effect\cite{LFZimp}.

An interesting phenomenon that appears due to inhomogeneous superconductivity is the Tomasch effect
(TE), which is a consequence of quasiparticle interference (QPI) due to scattering on a nonuniform
energy-gap.\cite{TE1, TE2, TE3, TE4}  The underlying process consists in a quasiparticle interacting
with, and being condensed into, the sea of Cooper pairs leaving behind a different but degenerate
quasiparticle.\cite{TE3, TE4}  As a result, Tomasch oscillations appear as periodic oscillations in
the density of states (DOS) in superconducting junctions, at energies larger than the
superconducting gap.\cite{TE1, TE2, TEE, vortex6}  These oscillations could be further
enhanced when considering layered structures formed by intercalating successive metallic and
superconducting regions. In this case the order parameter is by design non-homogeneous.
Theoretically, these oscillations were studied in multi-layer structures by using Green's functions
methods like Gor'kov equations.\cite{TEDOS1, TEDOS2} However, the electronic structure under TE
has not been unveiled because a fully self-consistent numerical calculation is required in order to
obtain the coherence between quasiparticle states.

QPIs should also be observed in unconventional superconductors\cite{QPI1, QPI2, QPI3}, where the interference should be more pronounced due to the intrinsic inhomogeneous nature of the superconducting order parameter.  Theoretically, QPI due to a local superconducting order parameter suppression in large unconventional samples was demonstrated in Ref.~[\onlinecite{zhu2006}].  However, in nanoscale structures, quantum confinement modifies the symmetries of the electronic states.  Thus, the signatures of QPI in nanoscale superconductors, although of same nature, could be different in manifestation in large samples but with intrinsic inhomogeneities~\cite{zhu2006}.

In this paper, we investigate the electronic structure of clean nanoscale superconductors by solving the
Bogoliubov-de Gennes (BdG) equations self-consistently.  We focus on the TE which appears
above the superconducting gap.  High precision energy excitation spectra are needed in order to see
the effect of the QPI processes clearly.  Two geometries, nanobelts and nanowires, are
used as typical examples in order to unveil the properties of TE resulting from two- and three- subband QPIs.
The importance of the sample symmetry is discussed. We find that even in the presence of weak disorder, the Tomasch is robust.

It is important to keep in mind that the mean-field BdG approach has limited validity when
describing superconducting nanobelts and nanowires with diameters down to $\unit[10]{nm}$.\cite{FLUC1, FLUC2,
FLUC3, FLUC4}  For such nano-scale
samples, fluctuations might play an important role, but are totally neglected by a mean-field
method.  Moreover, the quasi-particles in the Landau-Fermi liquid theory are only well defined near
the Fermi level.  Therefore, the discussion and results presented in this paper are valid for larger
samples where these effects are not significant.

The paper is organized as follows.  In Sec.~\ref{sec:2}, we first investigate TE of two-subband QPI in
nanobelts.  The two-dimensional BdG equations are outlined for nanobelts in SubSec.~\ref{sec:2} A.
Properties of the electronic structures under the TE of two-subband QPI are presented in SubSec.~\ref{sec:2}~B.  A
description based on
a reduced $4\times4$ BdG matrix is next introduced in SubSec.~\ref{sec:2}~C in order to explain the
properties of TE as due to two-subband QPI.  A possible observable effect, modulated wave patterns in LDOS,
induced by TE of two-subband QPI is discussed in SubSec.~\ref{sec:2}~D.  Signatures of QPI under the influence of weak impurities are discussed in SubSec.~\ref{sec:2}~E.
Next, we investigate TE of three-subband QPI in nanowires in
Sec.~\ref{sec:3}, where the three-dimensional BdG equations are solved for nanowires in
SubSec.~\ref{sec:3}~A. The electronic
structure under the influence of TE of three-subband QPI and their symmetry dependent properties are
presented in SubSec.~\ref{sec:3}~B. Finally, our conclusions are summarized in Sec.~\ref{sec:4}.

\section{Tomasch effect in superconducting nanobelts}\label{sec:2}
\subsection{Bogoliubov-de Gennes equations for two-dimensional nanobelts}
For a conventional superconductor in the clean limit, the BdG equations in the absence of a magnetic field can be written as
\begin{eqnarray}
\label{BdG}
\begin{pmatrix}
H_e & \Delta(\vec{r}) \\
\Delta(\vec{r})^\ast & -H_e^\ast
\end{pmatrix}
\binom{u_n(\vec{r})}{v_n(\vec{r})}
=E_n\binom{u_n(\vec{r})}{v_n(\vec{r})},
\end{eqnarray}
where $H_e=-(\hbar\nabla)^2/2m-E_F$ is the single-electron Hamiltonian with $E_F$ the Fermi energy,
$u_n$($v_n$) are electron(hole)-like quasiparticle eigen-wavefunctions and $E_n$ are the quasiparticle eigen-energies.
The $u_n$($v_n$) obey the normalization condition
\begin{equation}\label{Norm}
\int {(|u_n(\vec{r})|^2+|v_n(\vec{r})|^2)d\vec{r}}=1.
\end{equation}
The superconducting order parameter is determined self-consistently from the eigen-wavefunctions and eigen-energies:
\begin{equation}\label{OP}
\Delta(\overrightarrow{r})=g\sum\limits_{E_n<E_c}u_n(\vec{r}) v^\ast_n(\vec{r})[1-2f(E_n)],
\end{equation}
where $g$ is the coupling constant, $E_c$ is the cutoff energy, and $f(E_n)=[1+\exp(E_n/k_BT)]^{-1}$ is the Fermi distribution function, where $T$ is the temperature.  The core part of Eq. (\ref{OP}) is the pair amplitude which is defined as
\begin{equation}\label{paircorr}
D_n(\vec{r})=u_n(\vec{r}) v^\ast_n(\vec{r}).
\end{equation}
The pair amplitude is the key parameter that shows the coupling between electron-like and hole-like quasiparticles for each Bogoliubov quasiparticle.

In this section, we consider a two-dimensional nanobelt.  The width is $W$ in the transverse
direction, $x$, and, because of confinement, Dirichlet boundary conditions are used at the surface
(i.e. $u_n|_{x=0}=u_n|_{x=W}=0$, $v_n|_{x=0}=v_n|_{x=W}=0$).  We consider periodic boundary
conditions along the $y$ direction, with a unit cell of length $L$.  The length is set to be large
enough in order to guarantee that physical properties are $L$-independent.

In order to solve the BdG equations (\ref{BdG}-\ref{OP}) numerically, we expand $u_n$($v_n$) as
\begin{equation}\label{expand}
\binom{u_n(\vec{r})}{v_n(\vec{r})}=\sum_{j\in\mathbb{N}^+,k}\varphi_{jk}(x,y)\binom{u^n_{jk}}{v^n_{jk}}
\end{equation}
where
\begin{equation}\label{Fourier}
\varphi_{jk}(x,y)=\sqrt{\frac{2}{W}}sin\left ( \frac{\pi jx}{W} \right )\frac{e^{iky}}{\sqrt{L}}
\end{equation}
are the eigenstates of the single-electron Schr\"odinger equation $H_e\phi_{jk}=\zeta_{jk}\phi_{jk}$ where the wave vector $k=2\pi m /L, m\in \mathbb{Z}$.  The expansion in Eq.(\ref{expand}) has to include all the states with energy in the range $-E_F<\zeta_{jk}<E_F+\epsilon$ in order to allow the emergence of the TE well above the the energy gap.  The energy $\epsilon$ is taken to be $5E_c$, which guarantees sufficient accuracy. We checked that higher $\epsilon$ does not change the results.

We remark that, for the chosen geometry, the order parameter depends only on the transverse variable $x$, i.e., $\Delta(\vec{r})=\Delta(x)$.  This implies no net momentum of the condensate in the $y$ direction and the quasiparticle amplitudes $(u_n,~v_n)^T$ are $k$-separated.  Then, the summation over $k$ in Eq.~(\ref{expand}) can be removed and the BdG equation (\ref{BdG}) is converted into a matrix equation for each $k$ whose contribution to $\Delta$ can be calculated independently from the other values of $k$.  This allows us to include millions of quasiparticle states allowing very high resolution in the energy dispersion which is necessary to observe clearly the Tomasch effect.

The local density of states (LDOS) is calculated as usual:
\begin{equation}\label{LDOS}
A(\vec{r},E)=\sum\limits_{n} [|u_n(\vec{r})|^2\delta(E-E_n) +|v_n(\vec{r})|^2\delta(E+E_n)],
\end{equation}
and the total density of states (DOS) is obtained as
\begin{equation}\label{DOS}
N(E)=\int A(\vec{r},E) d\vec{r}.
\end{equation}
The spectral weight is
\begin{equation}\label{sweight}
Z_n=\int |u_n(\vec{r})|^2d\vec{r},
\end{equation}
which represents the contribution of the electronic part of the wave function of a Bogoliubov quasiparticle state.

In this section, we set the microscopic parameters to be the same as those used in Refs.~\onlinecite{vortex3, vortex6}.  These parameters for bulk are the following: effective mass $m=2m_e$, $E_F=\unit[40]{meV}$, $E_c=\unit[3]{meV}$ and coupling constant $g$ is set so that the bulk gap at zero temperature is $\Delta_0=\unit[1.2]{meV}$, which yields $T_c\approx \unit[8.22]{K}$, $\xi_0={\hbar}v_F/(\pi\Delta_0)=\unit[14.7]{nm}$ and $k_F\xi_0=21.23$.  The prototype material can be, e.g., $NbSe_2$.\cite{vortex3, vortex6}  For nanobelts, the mean electron density $n_e$ is kept to the value obtained when $W,L \rightarrow \infty $ by using an effective $E_F$, where $n_e=\frac{2}{S}\sum_{n} \int \left \{ |u_n|^2f(E_n)+|v_n|^2[1-f(E_n)]\right \}d\vec{r}$ and $S=W\times L$ is the area of the unit cell. All the calculations are performed at zero temperature.

\subsection{Tomasch effect from two-subband quasiparticle interference}

%
\begin{figure*}
\includegraphics[width=18cm,height=11cm]{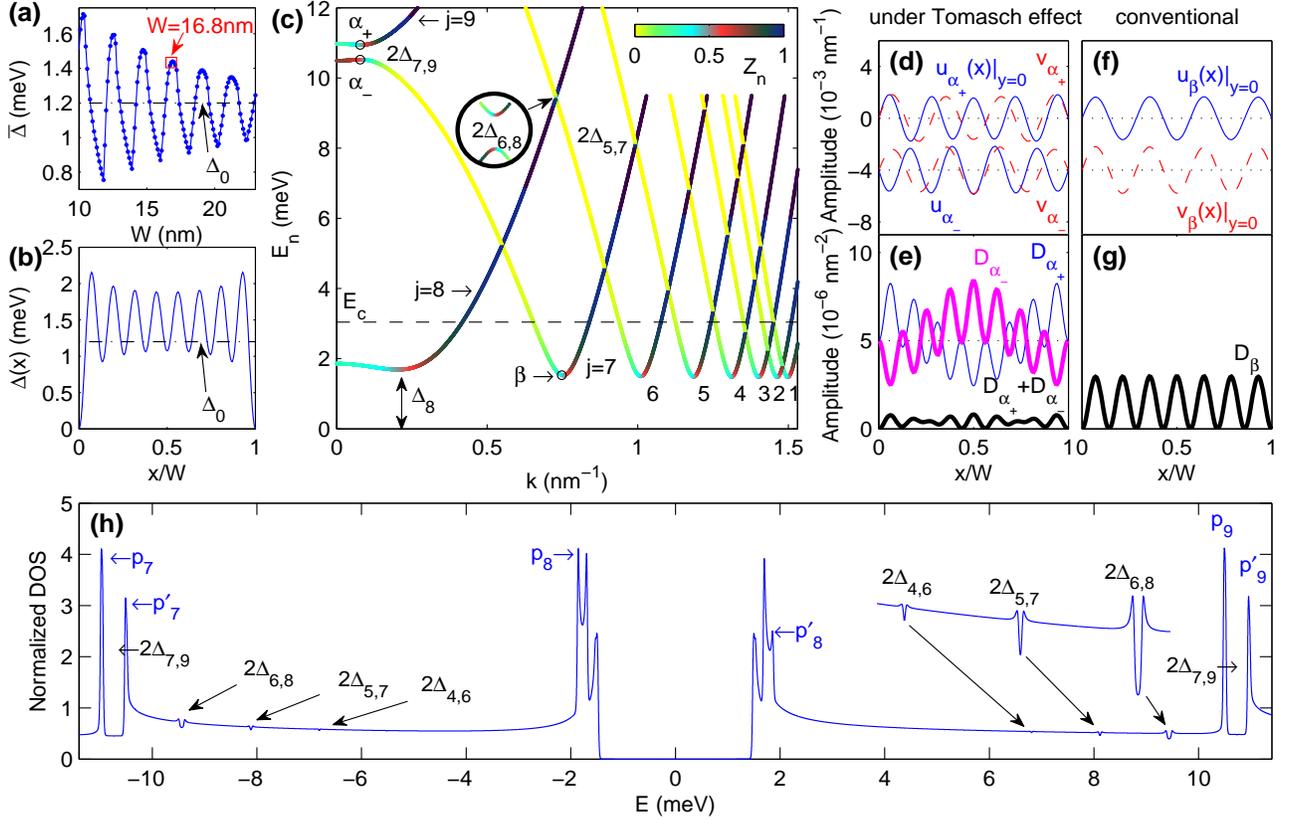}
\caption{(Color online) Nanobelt in the clean limit.  (a) Spatially averaged $\bar{\Delta}$ as a
function of width $W$.  In the following we take, $W=\unit[16.8]{nm}$ (marked by an open square), which is
in the resonant regime.  (b) Order parameter, $|\Delta(x)|$, as a function of $x$.  The horizontal
dashed line indicates the bulk value of the order parameter, $\Delta_0$, in panel (a) and (b).  (c)
Quasiparticle excitation spectrum as a function of positive longitudinal wave vector, $k$, for the
relevant single-electron subband $j$. The spectral weight of the states is indicated by color.  The
horizontal dashed line indicates the cutoff energy $E_c$.  $\Delta_j$ is the superconducting energy
gap for states of subband $j$ and $2\Delta_{jj'}$ is the energy gap between the states of subbands
$j$ and $j'$ appearing due to the Tomasch effect.  The electron and hole amplitudes, $u$ and $v$,
(at $y=0$ as a function of $x$) of states $\alpha_+$ and $\alpha_-$ [see panel (c)], which are
under the influence of the Tomasch effect, are shown in panels (d).  Their pair amplitudes,
$D_{\alpha_+}$ and $D_{\alpha_-}$, and the total one, $D_{\alpha_+}+D_{\alpha_-}$, are shown in
panel (e).  Note that $D_{\alpha_+}$ and $D_{\alpha_-}$ are shifted up for clarity.  In panels (f)
and (g), we show $u$, $v$ and $D$ as a function of $x$ for the state $\beta$ [see panel (c)],
respectively.  (h) The corresponding density of states.  Peaks $p_j$ are due to the contributions
from the states of the bottom of the subbands $j$, while accompanying peaks $p\prime_{j}$ are due to the
formation of the gaps, i.e. electron-hole coupling.}\label{fig.1}
\end{figure*}
%

Results for the nanobelt with width $W=\unit[16.8]{nm}$ are presented in Fig.~\ref{fig.1}.  We first discuss its general properties which will be used as a reference later on when the system is under the influence of TE.

The width of the nanobelt is about $\xi_0$ such that the quantum size effect is significant.  The spatially averaged order parameter $\bar{\Delta}$ is shown in Fig.~\ref{fig.1}(a) as a function of the width $W$.  This nanobelt with width $W=\unit[16.8]{nm}$ is at resonance and $\bar{\Delta}$ is about $20\%$ higher than the bulk one, $\Delta_0$.  As seen from Fig.~\ref{fig.1}(b), the spatial distribution of the order parameter, $\Delta(x)$, is clearly enhanced for most of the x-values and shows strong wave-like oscillations.  This enhancement is induced by quasiparticle states from the bottom of subband $j=8$ which have a large spectral weight as seen from Fig.~\ref{fig.1}(c).  For this narrow nanobelt, only a few subbands ($j=1-8$) contribute to the order parameter and are distinguishable from each other in Fig.~\ref{fig.1}(c).

For a superconductor under quantum confinement and thus with inhomogeneous order parameter, a multi-gap structure was predicted.\cite{MultiGap, NAV}  The energy gap $\Delta_j$ of the subband $j$ is defined by the lowest energy of that subband.  As seen from Fig.~\ref{fig.1}(c), the energy gaps $\Delta_{1},...,\Delta_{7}$ have almost the same value but are lower than $\Delta_8$.  This feature can be seen clearly in the corresponding DOS, in Fig.~\ref{fig.1}(h), where three significant pairs of coherence peaks form around the Fermi level.  The lowest energy pair of peaks are less important and are due to the contributions of quasiparticles of subbands $j=1-7$.  The second lowest pair of peaks are more significant and most of the contributions are given by quasiparticle states of the subband $j=8$ around $k=\unit[0.2]{nm^{-1}}$ where it reaches its local minimum, as shown in Fig.~\ref{fig.1}(c).  It is worth noting that these two pairs of peaks show electron-hole symmetry in the DOS, whereas the third lowest pair of peaks do not.  The latter ones are due to the contribution from states of subband $j=8$ around $k=0$.  The more significant peak in negative bias is due to: 1) a large number of hole quasiparticle states at the bottom of subband $j=8$, and 2) the electron-hole coupling asymmetry due to the higher energy level where the Bogoliubov quasiparticle states are formed by the larger weight of the hole component.  This can be seen from Fig.~\ref{fig.1}(c) where the spectral weight $Z_n$ represents the color(shade) of the lines.

In an isotropic superconductor, Bogoliubov quasiparticles are well defined only for energies close to the superconducting gap $\Delta$.  For such states, the electron and hole components are of the same weight, which maximizes the amplitude of the pair amplitudes that generate the order parameter.  With increasing energy, Bogoliubov quasiparticles decompose into dominant electron and hole components, accompanied by a dramatically decreasing pair amplitude.  Finally, they decompose into separate electron or hole quasiparticles belonging to the normal state.

For a conventional Bogoliubov quasiparticle, which is well formed for energies near the gap, the electron and hole components belong always to the same subbands $j$, i.e.  $j-j$ coupling.  As an example, we show more details of the lowest energy state $\beta$ of subband $j=7$ [marked by an open circle in Fig.~\ref{fig.1}(c)] in Figs.~\ref{fig.1}(f) and (g).  The electron and hole amplitudes at $y=0$, i.e. $u_\beta|_{y=0}$ and $v_\beta|_{y=0}$, as a function of $x$ are shown in Fig.~\ref{fig.1}(f). $v_\beta$ are shifted down for clarity.  It can be seen that both components are in phase because they belong to the same subband $j=7$.  Thus, the pair amplitude $D_{\beta}$ shown in Fig.~\ref{fig.1}(g) as a function of $x$, is always positive and shows a regular wave-like pattern with a constant envelope.

We next discuss the influence of the TE on the electronic structure.  First, we find that energy gaps are unexpectedly opened between electron and hole quasiparticle states even well above the Fermi level.  As seen from Fig.~\ref{fig.1}(c), the most pronounced energy gap above $E_c$ is generated by states $\alpha_{+}$ and $\alpha_{-}$ (marked by open circles) where subbands $j=7$ and $j=9$ were supposed to cross each other in case of a homogeneous superconductor.  Here, the states $\alpha_{\pm}$ have the same k-value and they are chosen because they have the minimal gap $2\Delta_{7,9}$ between the two subbands, i.e., $E_{\alpha_+}-E_{\alpha_-}=2\Delta_{7,9}$.  The energy of state $\alpha_{+}$ is higher than the state of $\alpha_{-}$.  As there are only two quasiparticles that take part in the QPI, we refer to this effect as the Tomasch effect determined by two-subband quasiparticles interference processes.

Second, we find that particle-hole mixing (Bogoliubov quasiparticles) is always significant for states under the influence of the TE.  This can be seen from the spectral weight of the related quasiparticle states in Fig.~\ref{fig.1}(c). Their color changes gradually for both the upper and lower energy bands that are associated with this gap.  Furthermore, the electron and hole wave-functions of the states under the TE have a different structure because they belong to different subbands, $j$ and $j'$, i.e. $j-j'$ coupling.  To see this characteristic, we show electron and hole amplitudes at $y=0$, $u|_{y=0}$ and $v|_{y=0}$, of the states $\alpha_{\pm}$ as a function of $x$ in Fig.~\ref{fig.1}(d).  Note that the amplitudes of the state $\alpha_{-}$ are shifted down for clarity.  For both states, $\alpha_{\pm}$, the spectral weights of the electron components are the same as those of the hole components, which resembles the conventional Bogoliubov quasiparticle state $\beta$.  However, their electron and hole wave-functions show a phase shift because the electron components, $u_{\alpha_{\pm}}$, belong to subband $j=9$ and the hole component, $v_{\alpha_{\pm}}$, to subband $j=7$.  This can be noticed by counting the numbers of nodes of the wave functions in Fig.~\ref{fig.1}(d).  The difference between these states is that $\alpha_{+}$ is the bonding(symmetric) combination of the electron and hole components while  $\alpha_{-}$ is the antibonding(anti-symmetric) combination, i.e., $|\alpha_\pm>=\pm |u_{j=9}>+|v_{j=7}>$, as seen from Fig.~\ref{fig.1}(d).  As a result, their pair amplitudes $D_{\alpha_{\pm}}$ as a function of $x$, shown in Fig.~\ref{fig.1}(e) are not positive-definite and exhibit a phase shift in space.

Third, the Bogoliubov quasiparticle states under the influence of TE do not  directly affect the order parameter.  As seen from Fig.~\ref{fig.1}(e), the pair amplitudes $D_{\alpha_{\pm}}$ are in anti-phase due to the nature of bonding and anti-bonding combinations.  Thus, their net pair amplitude $D_{net}=D_{\alpha_{+}}+D_{\alpha_{-}}$ almost cancels out as shown in Fig.~\ref{fig.1}(e).  It is worth noting that, as proven in the next subsection, the net pair amplitude of these states have nothing to do with the TE.  This contribution is always positive, while the superconducting order parameter is weakly affected by TE even if  the energy of this avoided crossing is below the cutoff energy $E_c$.

Fourth, TE results in BCS-like pseudo-gaps in the DOS, which  are symmetrically located on both positive and negative bias.  This is easy to understand because energy gaps are opened for the relevant crossing subbands where particle-hole mixing appears under the influence of the TE.  Typically, the gaps induced by the TE are smaller than the main superconducting gap. They will appear as pseudo-gaps because of the underlying background given by the other subbands which are not gapped. For example, as shown in Fig.~\ref{fig.1}(h), the largest gap induced by TE, $2\Delta_{7,9}$ is only about $\unit[0.42]{meV}$.  However, the effect of the gap resulting from the TE can be seen more clearly from the enhanced peaks appearing at the gap edge when the bottom of one subband touches the top of the other subband.  In Fig.~\ref{fig.1}(h), the gap $2\Delta_{7,9}$ is surrounded by the peaks $p_{j=7}$ and $p'_{j=7}$ in DOS for negative bias and by $p_{j=9}$ and $p'_{j=9}$ for positive bias.  The generation of the two pair of peaks are similar to the $p_{j=8}$, $p'_{j=8}$ except that they result from the top of the hole-like subband $j=7$ and from the bottom of the electron-like subband $j=9$.

For the chosen width, there are more gap structures induced by TE in the excitation spectrum [see Fig.~\ref{fig.1}(c)] and corresponding peaks in the DOS [see Fig.~\ref{fig.1}(h)].  For example, the gap $2\Delta_{6,8}$ appears for the coupling of states from subbands $j=6$ and $j=8$ at $E=\unit[9.4]{meV}$ but its influence on the DOS is weak.  We also find other gaps such as $2\Delta_{5,7}$, $2\Delta_{4,6}$ and $2\Delta_{3,5}$.

TE is a common effect in inhomogeneous superconductivity and is strongly related to the symmetry, parity and structure of the order parameter.  In the case of the clean limit, as we showed here, it is important to realize that avoided crossings exist only between electron and hole quasiparticle states of subbands $j$ and $j+2n$, where $n$ is an integer. This is because the order parameter has an even function with respect to $y=0$.  Similarly, an odd-functional order parameter would result in TE between states of subbands $j$ and $j+(2n-1)$.  In the arbitrary situation where the order parameter shows a random distribution due to strong disorder, TE should happen between all degenerate electron and hole quasiparticles.  All these properties can be explained by a reduced $4\times4$ BdG matrix as shown in the next subsection.

\subsection{$4\times4$ BdG matrix for the two-subband quasiparticle interference}
Due to the fact that only two subbands are involved in the TE of two-subband QPI, we find it can be qualitatively described by a reduced $4\times4$ BdG matrix where only the two Bogoliubov quasiparticles and their correlations are considered.

We start from the general BdG equations (\ref{BdG}) but only keep a hole state of subband $j$ and an electron state of subband $j'$ for a given wave vector $k$.  For the hole state, its energy and wave function are determined by the single-electron Schr\"odinger equation $H_e|j>=-\zeta_j|j>$.  Note that $-\zeta_j<0$ due to the hole excitation.  Again, the energy and wave function of the electron state is determined by $H_e|j'>=\zeta_{j'}|j'>$ where $E_{j'}>0$.  The orthogonal relation between the hole state $|j>$ and the electron state $|j'>$ yields $<j|j'>=\delta_{jj'}$.  For simplicity and fitting the case of previous subsection, $|j>$ and $|j'>$ are chosen as real and to generate the real order parameter $\Delta(\vec{r})$.  Then, the electron component of a Bogoliubov quasiparticle is $u_n=U_j|j>+U_{j'}|j'>$ and the hole component is $v_n=V_j|j>+V_{j'}|j'>$ where $U_j$ and $V_j$ are the component amplitude of subband $j$ for electron and hole, respectively.   The $4\times4$ BdG matrix reads:
\begin{eqnarray}\label{BdG44}
\begin{bmatrix}
-\zeta_j & 0 & \Delta_j & \Delta_{jj'}\\
0 & \zeta_{j'} & \Delta_{j'j} & \Delta_{j'}\\
\Delta_j & \Delta_{jj'} & \zeta_j & 0 \\
\Delta_{j'j} & \Delta_{j'} & 0 & -\zeta_{j'}
\end{bmatrix}
\begin{bmatrix}
U_{j}^{n}\\
U_{j'}^{n}\\
V_{j}^{n}\\
V_{j'}^{n}
\end{bmatrix}
=E_n
\begin{bmatrix}
U_{j}^{n}\\
U_{j'}^{n}\\
V_{j}^{n}\\
V_{j'}^{n}
\end{bmatrix}
\end{eqnarray}
with the matrix elements $\Delta_{j}=<j|\Delta(\vec{r})|j>$ and $\Delta_{jj'}=<j|\Delta(\vec{r})|j'>=\Delta_{j'j}$.  Note that the $\Delta_{jj'}$ is the exchange integral between the two states from different subbands.  In a homogeneous superconductor, the constant order parameter $\Delta(\vec{r})\equiv \Delta$ leading to the zero exchange integral, $\Delta_{jj'}=0$, results in the decomposition of Eqs.~(\ref{BdG44}) into two sets of general $2\times 2$ BdG matrices for the two states respectively.  Thus, there is no TE for this case.

For an inhomogeneous superconductor with a perturbation in the order parameter $\Delta(\vec{r})=\Delta+\delta\Delta(\vec{r})$, the matrix elements are
\begin{equation}\label{element3}
\Delta_{j}=<j|\Delta+\delta\Delta(\vec{r})|j>\approx\Delta,
\end{equation}
and
\begin{equation}\label{element4}
\Delta_{jj'}=<j|\Delta+\delta\Delta(\vec{r})|j'>=<j|\delta\Delta(\vec{r})|j'>=\Delta_{j'j}\neq0.
\end{equation}
Note that the non-zero exchange integral $\Delta_{jj'}\neq0$ in this case is responsible for the TE.

The TE of QPI reaches its maximum when states of two subbands are degenerate in energy, i.e., $\zeta_j=\zeta_{j'}=\zeta$.  So Eqs.~(\ref{BdG44}) are written as
\begin{eqnarray}\label{BdG44TE}
\begin{bmatrix}
-\zeta & 0 & \Delta & \Delta_{jj'}\\
0 & \zeta & \Delta_{jj'} & \Delta\\
\Delta & \Delta_{jj'} & \zeta & 0 \\
\Delta_{jj'} & \Delta & 0 & -\zeta
\end{bmatrix}
\begin{bmatrix}
U_{j}^{n}\\
U_{j'}^{n}\\
V_{j}^{n}\\
V_{j'}^{n}
\end{bmatrix}
=E_n
\begin{bmatrix}
U_{j}^{n}\\
U_{j'}^{n}\\
V_{j}^{n}\\
V_{j'}^{n}.
\end{bmatrix}
\end{eqnarray}
The eigenvalues and eigenstates of matrix (\ref{BdG44TE}) are exactly solvable and the four eigenvalues are
\begin{equation}\label{eigend}
\pm E_\pm=\pm \varepsilon\pm \Delta_{jj'}
\end{equation}
where $\varepsilon=\sqrt{\zeta^2+\Delta^2}$ is the quasiparticle excitation energy of the isotropic superconducting gap $\Delta$.  The gap induced by TE, $\Delta_{TE}$, is the energy difference between the two positive eigenvalues:
\begin{equation}\label{gap}
\Delta_{TE}=
\begin{cases}
2\Delta_{jj'} & \text{ if } \Delta_{jj'}<\varepsilon \\
2\varepsilon & \text{ if } \Delta_{jj'}\geqslant \varepsilon.
\end{cases}
\end{equation}

Typically, the exchange integral $\Delta_{jj'}$ is smaller than the excitation energy gap $\varepsilon$.  As a result, $\Delta_{TE}=2\Delta_{jj'}$ and that is why we labeled the gaps as $2\Delta_{j,j'}$ in Fig.~\ref{fig.1}.

When the exchange integral is positive, i.e., $0<\Delta_{jj'}<\varepsilon$, the eigenvalues sorted by their values are $\begin{pmatrix}-E_+ & -E_- & E_- & E_+ \end{pmatrix}$ and their corresponding eigenstates are
\begin{equation}\label{eigenv}
\begin{pmatrix}
U_j^n\\
U_{j'}^n \\
V_j^n\\
V_{j'}^n
\end{pmatrix}
=
\begin{pmatrix}
A & A & B & B\\
B & -B & -A & A\\
-B & -B & A & A\\
-A & A & -B & B
\end{pmatrix}
\end{equation}
where
\begin{equation}\label{UV}
\begin{matrix}
A=\frac{1}{2}\left ( 1+\frac{\zeta}{\varepsilon} \right )^{\frac{1}{2}}\\
B=\frac{1}{2}\left ( 1-\frac{\zeta}{\varepsilon} \right )^{\frac{1}{2}}.
\end{matrix}
\end{equation}
Normalization is chosen to satisfy Eq.~(\ref{Norm}), i.e. $2(A^2+B^2)=1$.  For both eigenstates with positive eigenvalue $E_+$ and $E_-$, their spectral weights $Z_{E_{\pm}}=0.5$ indicate that Bogoliubov quasiparticles are well formed by the coupling between the electron and hole subbands.  The difference between the two states are the bonding and anti-bonding combinations of the electron and hole components.

It is interesting to realize that the $A$ and $B$ are $\Delta_{jj'}$-independent and $\sqrt{2}(A~B)^T$ is the eigenstate of the positive eigenvalue of the general $2\times2$ BdG equations, i.e.,
\begin{equation}\label{BdG22}
\begin{pmatrix}
\zeta & \Delta \\
\Delta & -\zeta
\end{pmatrix}
\begin{pmatrix}
\sqrt{2}A \\
\sqrt{2}B
\end{pmatrix}
=\varepsilon
\begin{pmatrix}
\sqrt{2}A \\
\sqrt{2}B
\end{pmatrix}.
\end{equation}
where the eigen-energy is $\varepsilon$ and the $\sqrt{2}$ is introduced to satisfy the normalization condition, Eq.(\ref{Norm}).  It turns out that the total pair amplitude of the states $E_+$ and $E_-$ are the same as the one without TE, i.e.,
\begin{equation}\label{Dpm}
D_{E_+}+D_{E_-}=2AB(|j>^2+|j'>^2)=D_j+D_{j'}.
\end{equation}

Finally, we have to mention that the exchange integral $\Delta_{jj'}$ is sensitive to the symmetry, parity and spatial variation of the order parameter $\Delta(\vec{r})$.  For the nanobelt in the clean limit, $\Delta(\vec{r})$ has a spatial distribution with even-parity with respect to $y=0$.  The exchange integral is exactly zero when both states $|j>$ and $|j'>$ have different parity.  This is the reason why TE only appears between electron and hole quasiparticle states of subbands $j$ and $j+2n$ which have the same parity, resulting in a possible non-zero exchange integral $\Delta_{jj'}$.

\subsection{Modulated waves in the local density of states due to Tomasch Effect}

%
\begin{figure}
\includegraphics[width=\columnwidth]{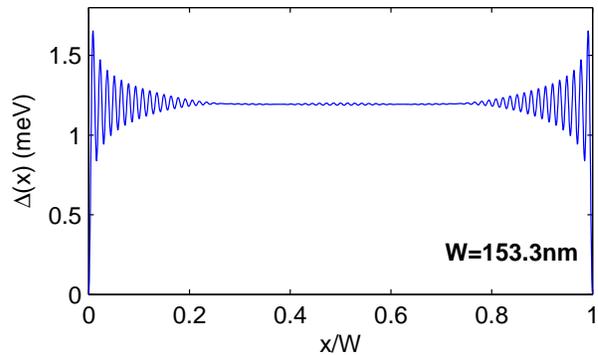}
\caption{(Color online) Spatial order parameter distribution $|\Delta(x)|$ for a sample with
$W=\unit[153.3]{nm}$.  Note that the order parameter converges to $\Delta_0$ in the bulk.}\label{fig.12}
\end{figure}
%

Previously, we introduced the properties of the TE of two-subband QPIs for a narrow sample.  However, the mean field BdG theory is of limited validity in such a case  due to the increasing importance of phase fluctuations and, moreover, quasiparticles are not well defined far above the Fermi level.  In this subsection, we investigate the TE in wider samples in order to avoid these issues.  The results of this subsection show that: 1) properties obtained previously are still valid, and 2) the TE results in a modulated wave structure in the local density of states, which should be observable in experiments.

%
\begin{figure*}
\includegraphics[width=14cm,height=10cm]{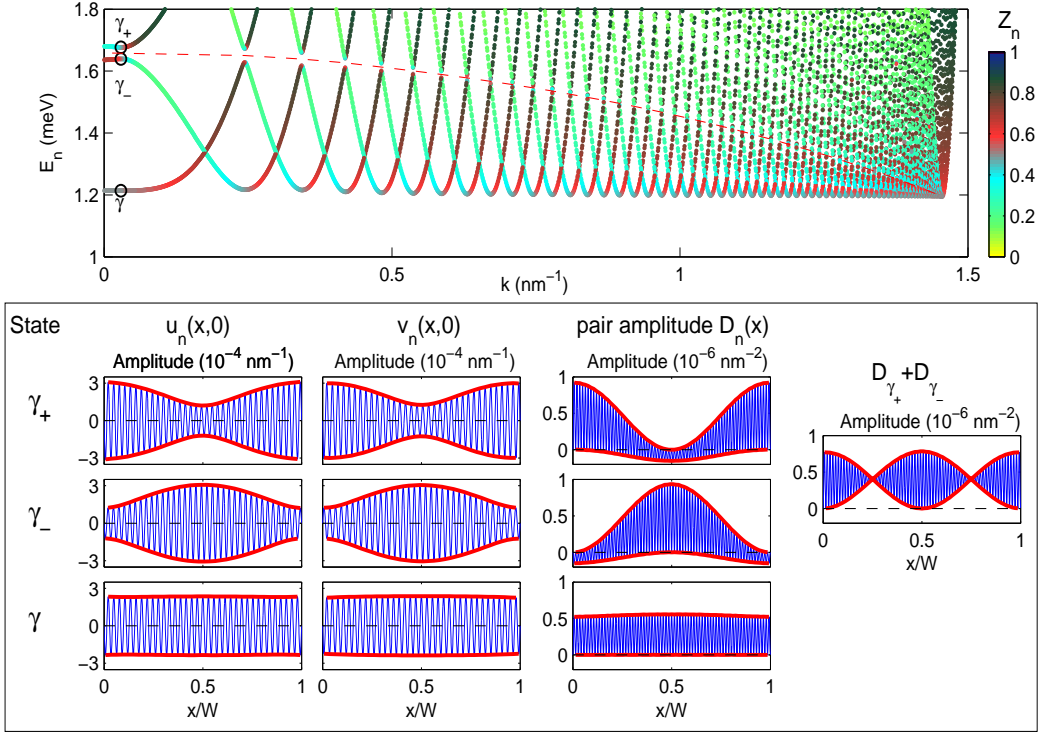}
\caption{(Color online) (Upper panel) Quasiparticle excitation spectrum as a function of positive
longitudinal wave vector, $k_z$, for energies below $\unit[1.8]{meV}$ for a nanobelt of width $W=\unit[153.3]{nm}$.
The spectral weight of the quasiparticle states are indicated by color.  The three quasiparticle
states $\gamma_+$, $\gamma_-$ and $\gamma$ marked by open circle are chosen for showing the electron
and hole amplitudes, $u$ and $v$, as a function of $x$ at $y=0$ and their pair amplitudes $D$ in
the lower panels. The red thick lines are the envelopes of the curves.  The quasiparticle states
$\gamma_+$ and $\gamma_-$ are under the influence of the Tomasch effect and show a phase
difference between their $u$ and $v$ components.  The net pair amplitude of the two states
$D_{\gamma_+}+D_{\gamma_-}$ is shown in the right-most panel.}\label{fig.13}
\end{figure*}
%

As an example, we present results for a nanobelt with width $W=\unit[153.3]{nm}$, which is more than $10\xi_0$.  The spatial distribution of the order parameter $\Delta(x)$ is shown in Fig.~\ref{fig.12}.  It can be seen that the order parameter shows Friedel-like oscillations at both edges but it converges to its bulk value $\Delta_0$ far away from the edges.  The flat order parameter in the central area suppresses the TE.  Fortunately, the energy gaps induced by TE can still be seen clearly in the corresponding quasiparticle excitation spectrum in the upper panel of Fig.~\ref{fig.13}.  Here, we focus on the gaps at the intersection between states of subbands $j$ and $j+2$, which are indicated by a dashed curve.  Due to the smaller energy difference between the adjacent subbands in the wider sample, the gaps appear at energies close to the superconducting gap energy $\Delta_0$ and far below the cutoff energy $E_c$, where the quasiparticles are well defined.

The TE exhibits all the properties which have been introduced previously except that the quasiparticle states generate considerable net pair amplitude, contributing to the order parameter.  To show this feature, we present the electron and hole amplitudes, $u$ and $v$, and their pair amplitude, $D$, of selected quasiparticle states $\gamma_+$ and $\gamma_-$ (marked by open circle in the upper panel) as a function of $x$ in the lower panels of Fig.~\ref{fig.13}.  Both states have the same wave vector $k$ and are separated by an energy gap due to the influence of TE.  It can be seen that $u$ and $v$ exhibit rapid oscillations with slowly varying envelopes. The envelopes show modulated wave structures due to the combination of states of subbands $j$ and $j+2$.  The difference in envelope for states $\gamma_+$ and $\gamma_-$ are due to the bonding and antibonding combinations of the two single-electron wave functions, respectively.  Meanwhile, the phase shift between the $u$ and $v$ components leads to more complex pattern in their pair amplitude.  Finally, the net pair amplitude $D_{\gamma_+}+D_{\gamma_-}$ (shown in the lower panel of Fig.~\ref{fig.13}) is large and positive, showing a modulated wave structure.

As a reference, we also present the electron amplitude, $u$, hole amplitude, $v$, and its pair amplitude, $D$, of a conventional Bogoliubov quasiparticle state $\gamma$ (also marked by open circle in the upper panel) as a function of $x$ in the lower panels of Fig.~\ref{fig.13}.  Because $u$ and $v$ belong to the same subband $j$, they are in phase leading to a positive pair amplitude with a flat envelope.

%
\begin{figure}
\includegraphics[width=\columnwidth]{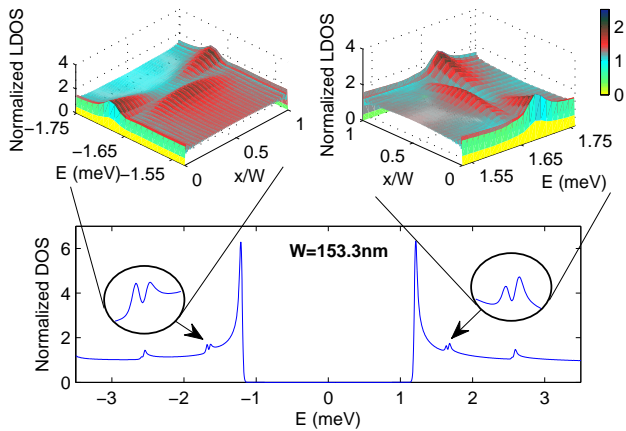}
\caption{(Color online) DOS and relevant LDOS for the sample with $W=\unit[153.3]{nm}$.  The DOS is shown in
the lower panel.  The oscillatory structures induced by the Tomasch effect are magnified in the two
insets.  The corresponding LDOS $A(x,E)$ along the transversal direction are shown in the upper
panels.}\label{fig.14}
\end{figure}
%

The states $\gamma_+$ and $\gamma_-$, which are under the influence of TE, induce peaks in the DOS and modulated wave structures in the LDOS.  Fig.~\ref{fig.14} show the corresponding DOS in the lower panel and the LDOS under the influence of TE in the upper panel.  In the DOS, the peaks induced by states $\gamma_\pm$ sit at the symmetrical bias energy $E=\pm \unit[1.65]{meV}$.  The insets magnify the relevant areas.  In the insets, the outer peaks are induced by the state $\gamma_+$ while the inner peaks are induced by the state $\gamma_-$.  The LDOS shows very different patterns at these two energies, which can be seen from the upper panels.  For the outer peaks in DOS, the LDOS is enhanced at the edge whereas, for the inner peaks, it is enhanced at the center.  The envelope of LDOS varies slowly as a function of $x$.  Therefore, it may be easily detected by STM.

\subsection{Discussion on the signature of the Tomasch effect in the presence of disorder}

%
\begin{figure}
\includegraphics[width=\columnwidth]{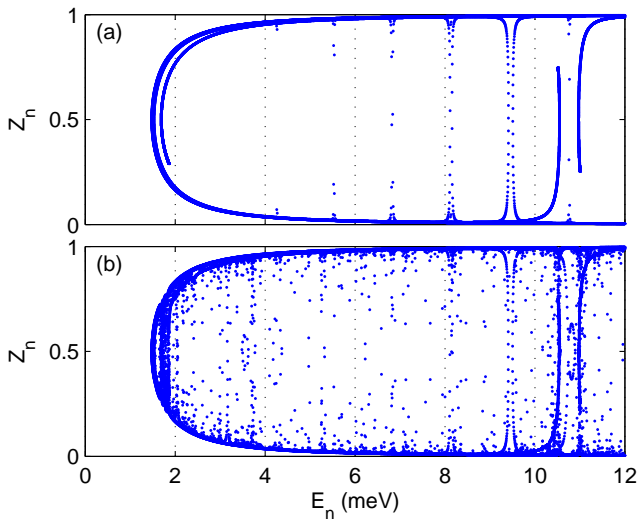}
\caption{(Color online) Spectral weight of quasiparticles, $Z_n$, as a function of energy of states, $E_n$, for the nanobelt of width $W=\unit[16.8]{nm}$ (a) in the clean limit (U=0) and (b) under the influence of random impurities.}\label{fig.15}
\end{figure}
%

%
\begin{figure}
\includegraphics[width=\columnwidth]{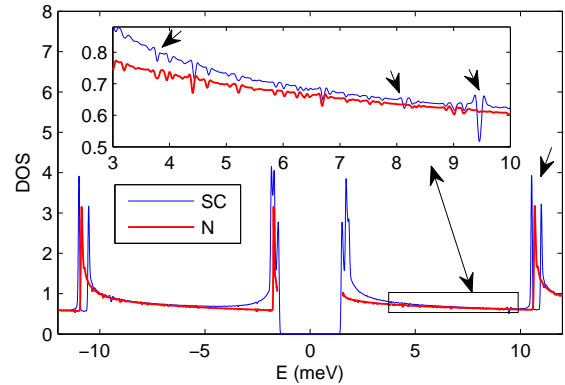}
\caption{(Color online) DOS of the superconducting (blue) and normal (red) states for the nanobelt of width $W=\unit[16.8]{nm}$ under the influence of the random impurities.  The inset shows details of the DOS between $E=\unit[3]{meV}$ and $\unit[10]{meV}$.  The arrows indicate signatures of the TE.  Note that the normal DOS is mapped on a new energy scale, in order to match the oscillations in the DOS induced by the impurities.  The mapping is done by rescaling, $E_{new}=\sqrt{E_{old}^2+|\tilde{\Delta}|^2}$, where $E_{new}$ and $E_{old}$ are the new and old energies, and $\tilde{\Delta}$ is a fitting parameter whose value is close to the mean amplitude of the order parameter $|\bar{\Delta}|$.}\label{fig.16}
\end{figure}
%

Previously, only the TE in clean superconductors were considered.  However, in all experiments, superconductors are rather "dirty", where additional scattering processes of quasiparticles appear due to surface roughness, impurities, substrate and so on.  These factors will broaden the single-electron levels,  modify electronic wave functions, reduce the electron mean free path and, thus,  lead to noise and modifications in DOS and LDOS.   In this subsection, we study TE under such additional scattering processes by introducing a random distribution of weak impurities.  The results of this subsection show that :1) many more gaps induced by TE are opened in this case, 2) one can still recover the dominant gaps seen the clean limit by comparing the DOS of superconducting and normal states,  3) the LDOS shows more complex scattering patterns but follows the same bonding and antibonding combination of involved quasiparticle wave functions.

The random impurities are introduced by using the potential $U(\overrightarrow{r})$ in the single-electron Hamiltonian defined Eq.~(\ref{BdG}), i.e.  $H_e=-(\hbar\nabla)^2/2m+U(\overrightarrow{r})-E_F$.  As a result, the order parameter, $\Delta(\overrightarrow{r})$, is no longer longitudinally independent, i.e.  $\Delta(\vec{r})=\Delta(x,y)$.  Following the numerical approach introduced in Sec.~\ref{sec:2}~A, we take a periodic unit cell with length $L$ and width $W$ and expand the electron(hole)-like quasiparticle wavefunctions $u_n(v_n)$ by using Eq.~(\ref{Fourier}).  Using Bloch's theorem, we the BdG equations will separate for each reciprocal vector. By considering a large number of k points, we achieve a good resolution in the DOS in order to observe the TE.  In this subsection, we take $W=\unit[16.8]{nm}$ and $L=\unit[40]{nm}$.  Such nanobelt in the clean limit ($U=0$) has been introduced in Sec.~\ref{sec:2}~B.  The impurity potential profile is modeled by a random collection of symmetric Gaussian functions, $U(\overrightarrow{r})=\sum_{i} U_{i}exp[-(\overrightarrow{r}-\overrightarrow{r_i})^2/2\sigma^2]$ where $U_i$ is the amplitude, $\overrightarrow{r_i}$ is the location of the impurity and $\sigma=0.02$ represents the the spread of the potential (full width at $1/10$th of maximum is $\unit[0.86]{nm}$).  For the situation of weak impurities (disorder), we take $100$ impurities in the unit cell with random locations $r_i$ and random amplitude $U_i$ with a maximum $U_i^{max}=0.1E_F$.  After the BdG equations are solved self consistently, as described in the previous sections, the order parameter, $\Delta$, has almost the same distribution as the one with $U=0$.

We show in Fig.~\ref{fig.15} the spectral weight of quasiparticles $Z_n$ as a function of energy of states $E_n$ for the case in the clean limit and the case with impurities.  In clean bulk superconductors, particles and holes never mix at energiesl away from the superconducting gap, i.e. $Z_n=0~or~1$, for holes or electrons, respectively.  In the case of clean superconductor under quantum confinement, as seen from Fig.~\ref{fig.15}(a), particle-hole mixing indicates the emergence of TE due to the stripe-like inhomogeneity of the order parameter.  In the presence of the impurities, as shown in Fig.~\ref{fig.15}(b), TE appears for much wider range of energies due to the symmetry broken of the electronic wave functions.  It indicates that many more TE gaps are opened at the crossover of electron and hole subbands for more realistic situations. Nevertheless, for weak disorder, the stronger contribution is still observed at the same energies as obtained in the clean limit.

To find the signature of TE, we show the DOS of both superconducting and normal states in the presence of impurities in Fig.~\ref{fig.16}.  The noise-like oscillations in the DOS are imposed over the signature of TE.  After matching both oscillations in DOS by mapping the DOS of normal state to a rescale energy range, one can find TE signatures 1) where there are new oscillations in the superconducting DOS and, 2) where there are different oscillatory structures between DOS of superconducting and normal states.  In Fig.~\ref{fig.16}, these signatures are marked by arrows.   It is worth noting that the TE modifies the DOS on positive and negative biases symmetrically.

Finally, we have to mention that the LDOS under the influence of random impurities shows much more complex patterns.  However, the pattern still follows the bonding and antibonding combination of involved quasiparticle wave functions as described in the previous sections.

As the impurity strength increases, the band structure together with order parameter  become strongly affected. In this case, while the TE is also strongly enhanced, it becomes increasingly difficult to compare with results obtained in the clean limit. Depending on the particular impurity distribution, TE contributions to the DOS could be individually recognized but these are of different manifestation, when compared to the DOS modifications obtained in the clean limit.

\section{Tomasch effect in superconducting nanowires}\label{sec:3}

In this section, we consider superconducting nanowires with square and rectangular cross sections.  We find a new type of TE, i.e., TE induced by three-subband QPI. Its influence on the electronic structure and its dependence on the symmetry of the system will be discussed in the following sections.

\subsection{Bogoliubov-de Gennes theory for three-dimensional nanowires}

Here, we consider a three-dimensional superconducting nanowire with rectangular cross section (transverse dimensions $L_x$ and $L_y$).  Due to quantum confinement in the transverse directions, the Dirichlet boundary conditions are taken on the surface (i.e. $u_n(\overrightarrow{r})=v_n(\overrightarrow{r})=0,\; \vec{r} \in \partial S$).  Along the longitudinal direction $z$, we introduce a periodic computational unit cell with length $L_z$ where periodic boundary conditions are used.

Due to the fact that the order parameter is independent of the longitudinal direction, i.e. $\Delta(\vec{r})=\Delta(x,y)$, the electron-like and hole-like wave functions $u_n$ and $v_n$ can be expanded, for each longitudinal wave vector $k_z$, as
\begin{equation}\label{expand3d}
\binom{u_n(\vec{r})}{v_n(\vec{r})}=\frac{e^{ik_z z}}{\sqrt{L_z}} \sum_{j_x,j_y\in\mathbb{N}^+}\phi_{jk}(x,y)\binom{u^n_{j_xj_y}}{v^n_{j_xj_y}},
\end{equation}
where
\begin{equation}\label{Fourier3d}
\phi_{j_xj_y}(x,y)=\frac{2}{\sqrt{L_xL_y}}sin\left ( \frac{\pi j_xx}{L_x} \right )sin\left ( \frac{\pi j_yy}{L_y} \right ),
\end{equation}
are the eigenstates of the single-electron Schr\"odinger equation $H_e\phi_{j_xj_y}=\zeta_{j_xj_y}\phi_{j_xj_y}$.  The longitudinal momentum, $k_z$, satisfies the quantization condition, i.e. $k_z \cdot L_z=2\pi m, m\in \mathbb{Z}$.  Following the previous section, the expansion in Eq.~(\ref{expand3d}) includes the states with energies $-E_F<\zeta_{j_xj_y}<E_F+\varepsilon$ where $\varepsilon=5E_c$ is taken sufficiently large in order to guarantee the accuracy.

The pair amplitude $D_n(\vec{r})$ and spectral weight $Z_n$ for each state are calculated from the Eqs.~(\ref{paircorr}) and (\ref{sweight}), respectively.  The LDOS $A(\vec{r},E)$ and the DOS $N(E)$ are calculated from the Eqs.~(\ref{LDOS}) and (\ref{DOS}), respectively.

We use the same microscopic parameters as the one introduced in SubSec.~\ref{sec:2}~A for bulk $NbSe_2$.  The mean electron density $n_e=\frac{2}{V}\sum_{n} \int \left \{ |u_n|^2f(E_n)+|v_n|^2[1-f(E_n)]\right \}d\vec{r}$ for nanowires with $V=L_xL_yL_z$.  $n_e$ is kept to its bulk value obtained when $L_{x,y,z} \rightarrow \infty $.  All the calculations are performed at zero temperature.

\subsection{Tomasch effect due to three-subband quasiparticles interference}

%
\begin{figure*}
\includegraphics[width=15cm,height=9cm]{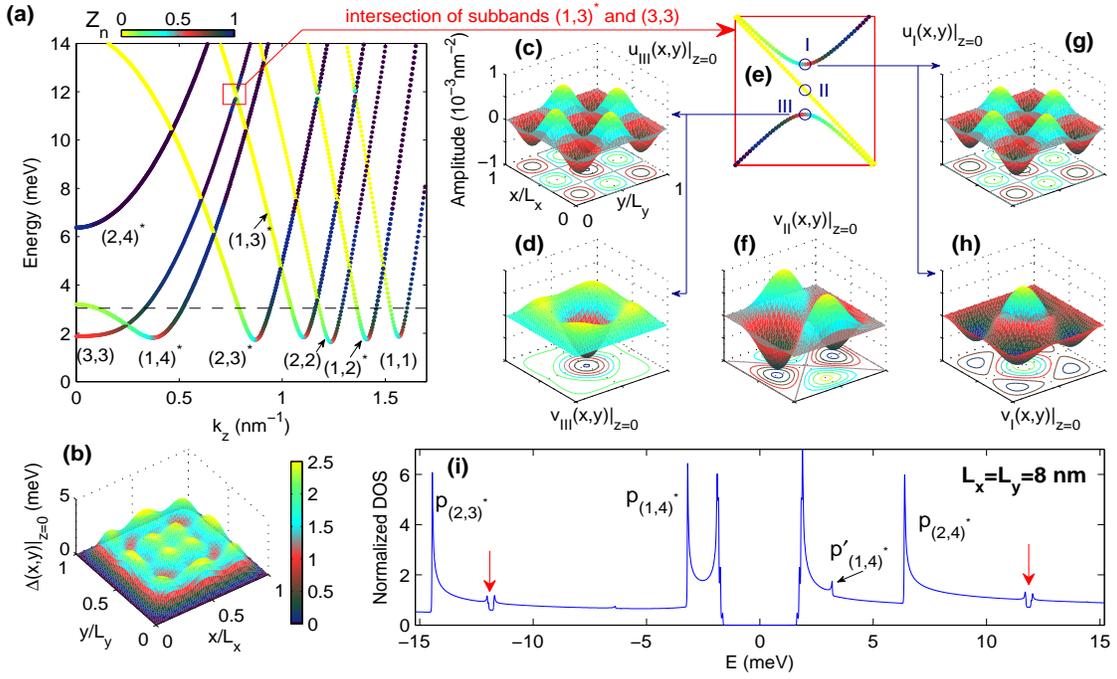}
\caption{(Color online) Results for the nanowire with a square cross section: $L_x=L_y=\unit[8]{nm}$.  (a)
Quasiparticle excitation spectrum as a function of positive longitudinal wave vector $k_z$ for the
relevant single-electron subbands $(j_x,j_y)$.  Note that the notation $(j_x,j_y)^*$ represents two
degenerate subbands $(j_x,j_y)$ and $(j_y,j_x)$.  The spectral weight of the quasiparticle states
is indicated by color. The excitation spectrum in the red rectangle is magnified in
panel (e) in order to show the influence of the Tomasch effect due to three-subband quasiparticle
interference. The three quasiparticle states $I$-$III$ are marked by open circle in panel (e) and
their spatial distribution of electron and hole amplitude, $u$ and $v$, at $z=0$ are shown in panels
(c, d, f-h). Note that the electron amplitude of the state $II$ is not shown because
$u_{II}(x,y)=0$. (b) The spatial distribution of the order parameter $\Delta(x,y)$.  (i) The
corresponding DOS. The gaps due to the Tomasch effect as determined by three-quasiparticle
interference processes are marked by red arrows.  Peaks $p_{(j_x,j_y)}$ are due to the contributions
from the states of the bottom of the subbands $(j_x,j_y)$, while accompanying peaks $p\prime_{(j_x,j_y)}$ are due to the
formation of the gaps, i.e. electron-hole coupling. }\label{fig.21}
\end{figure*}
%

We first show results in Fig.~\ref{fig.21} for a nanowire with square cross section ($L_x=L_y=\unit[8]{nm}$) where the geometry of the sample and the order parameter show $C_4$ symmetry. Fig.~\ref{fig.21}(a) shows the excitation spectrum as a function of the positive wave vector $k_z>0$.  All the subbands displayed in the panel are distinguishable and labeled by a set of quantum numbers $(j_x,j_y)$.  Note that the $(j_x,j_y)^*$ is a shorthand notation for the pairs $(j_x,j_y)$ and $(j_y,j_x)$ because of their overlap due to degeneracy.  The spectral weight $Z_n$ is marked by color for each quasiparticle state.

As seen from Fig.~\ref{fig.21}(a), the bottom of the subband $(3,3)$ lying below the cutoff energy $E_c$ results in a resonant and spatially inhomogeneous order parameter $\Delta(x,y)$, which is shown in Fig.~\ref{fig.21}(b).  $\Delta(x,y)$ shows $C_4$ symmetry due to the square cross section.  The symmetry of the system and the order parameter determine the properties of TE and its emergence.  For example, two sets of TE from two-subband QPI appear at the intersection of subbands $(1,2)^*$ and $(1,3)^*$, while TE from three-subband QPI appears at the intersection of subbands $(1,3)^*$ and $(3,3)$ or  $(1,3)^*$ and $(1,1)$, respectively.

We next investigate the most significant three-QPI appearing at the intersection of two hole quasiparticle states from subbands $(1,3)^*$ and one electron quasiparticle state from subband $(3,3)$.  This is indicated by the open red rectangle in Fig.~\ref{fig.21}(a) and the relevant three dispersion relations of the energy bands are amplified in Fig.~\ref{fig.21}(e).  It is clearly seen that the upper and lower energy bands exhibit a gap-like structure while the middle energy band crosses the gap diagonally.  The Bogoliubov quasiparticles are well formed for the states close to the bottom of the upper band and the top of the lower band.  Note that the state pertaining to the middle band are pure hole-like quasiparticles states with zero amplitude of the electron component.  This is an interesting phenomenon because if a gap opens between states from subbands $(1,3)$ and $(3,3)$, the other gap was supposed to be opened between states from subbands $(3,1)$ and $(3,3)$.  The reason is that the gap induced by the exchange integral only depends on the symmetry of the wave functions of the relevant states.

In fact, this interesting asymmetrical energy band structure is due to the symmetric and
anti-symmetric combinations of the two hole states from subbands $(1,3)^*$.  To see this, in
Figs.~\ref{fig.21}(c, d, f-h), we show the spatial distribution of the hole and the electron
amplitude, $u(x,y)$ and $v(x,y)$, of states $I$-$III$ marked by open circles in
Fig.~\ref{fig.21}(e).  The three states have the same wave vector $k_z$, chosen such that the gap
has a local minimum [see panel \ref{fig.21}(e)].  The electron components of the gapped states,
$u_I$ and $u_{III}$ [shown in panels \ref{fig.21}(c) and (g)],  have contributions only from
the electron state of subband $(3,3)$.  Therefore, they show the same pattern as $\phi_{3,3}$, which
is the eigenstate of the single-electron Schr\"odinger equation with quantum numbers $j_x=3$ and
$j_y=3$.  For the corresponding amplitude of hole components, $v_I$ and $v_{III}$ [shown in panels
\ref{fig.21}(d) and (h)], they have the same spatial distribution as the symmetric combination of
the two eigenstates, i.e., $\phi_{1,3}+\phi_{3,1}$, but with opposite sign for the two amplitudes I
and III.

Clearly, the Bogoliubov quasiparticle states, $I$ and $III$, are the bonding and anti-bonding
combinations of the electron and hole components.  The reason is that both the wave-functions and
the order parameter exhibit $C_4$ symmetry and, thus, result in non-zero exchange integrals, which
are responsible for TE and generate energy gaps.  The quasiparticle state $II$ does not take part in
the quasiparticle interference because its hole component, $v_{II}$, has $C_2$ symmetry,
($v_{II}(x,y)=-v_{II}(y,x)$), due to the anti-symmetric combination of the two hole eigenstates,
i.e., $\phi_{1,3}-\phi_{3,1}$, leading to a vanishing exchange integral.

We now conclude the appearance of TE due to three-subband QPI.  First, hole states from subbands
$(1,3)$ and $(3,1)$ are combined in order to generate symmetric and anti-symmetric states
but which are degenerate in energy.  Then, the symmetric combination couples with the
electron state from subband $(3,3)$ and forms Bogoliubov quasiparticle states, therefore inducing a
gap.  Finally, the energy and the wave-function of the anti-symmetric combination is
unaffected.

The process results in oscillations in the DOS, which are symmetrical in bias [see
Fig.~\ref{fig.21}(i)].  The oscillations induced by TE from the three-QPI are marked by arrows for
both positive and negative biases.  When comparing with the DOS of a nanobelt shown in
Fig.~\ref{fig.1}(h), we notice that there are less oscillations induced by TE.  The reason is that
TE emerges only in case of a non-zero exchange integral.  This becomes harder to achieve for a system
with two quantum numbers, $(j_x,j_y)$, because the condition has to be fulfilled by both .
%
%
\begin{figure}
\includegraphics[width=\columnwidth]{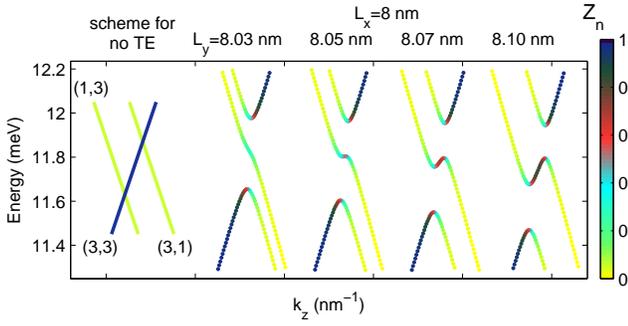}
\caption{(Color online) Symmetry-dependent Tomasch effect of three-subband quasiparticles interference for
states of subbands $(1,3)$, $(3,1)$ and $(3,3)$ for rectangular cross section $L_x<L_y$.  The
left-most panel shows the band dispersion for the three subbands without Tomasch effect.
The other four panels show the band dispersion versus longitudinal wave vector $k_z$ for
$L_x=\unit[8]{nm}$ and $L_y=\unit[8.03]{nm}$, $\unit[8.05]{nm}$, $\unit[8.07]{nm}$, $\unit[8.10]{nm}$, respectively. They are shifted
horizontally for clarity.  The spectral weight of the quasiparticle states is indicated by color in
all panels.}\label{fig.22}
\end{figure}
%
%
Next we will show that the TE of three-subband QPI in a nanowire depends strongly on the symmetry of the
electronic structures and the geometry of the sample.  For this purpose, we consider nanowires with
rectangular cross section, i.e. $L_x \neq L_y$ where the $C_4$ symmetry is broken and,
more importantly, results in the splitting of the subbands $(j_x,j_y)$ and $(j_y,j_x)$.  In
Fig.~\ref{fig.22}, we present the appearance of the TE due to three-subband QPI for states of hole subbands
$(1,3)$, $(3,1)$ and electron subband $(3,3)$ for nanowires with $L_x=\unit[8]{nm}$ and $L_y=\unit[8.03]{nm}$,
$\unit[8.05]{nm}$, $\unit[8.07]{nm}$ and $\unit[8.10]{nm}$, respectively.  The spectral weight of the relevant states is
indicated by color.  The three subbands without TE are shown in the most left panel. As mentioned
previously, the two hole subbands $(1,3)$ and $(3,1)$ split while the electron subband $(3,3)$
crosses them.  The states of subband $(3,1)$ have higher energy than those of subband $(1,3)$ for a
given $k_z$ when $L_y>L_x$.  It can be seen that the bottom of the highest energy subband shifts to
the right with increasing $L_y$, while the top of the lowest energy subband shifts to the left.
When comparing with the result for a square cross section shown in Fig.~\ref{fig.21}(e), we find
that the previously unaffected hole subband becomes and shows mixed electron-hole components,
signaling a coupling with the electron subband. Finally, the three-subband QPI converts into two sets
of two-subband QPI, appearing for states from subbands $(1,3)$ and $(3,3)$ and separately from
subbands $(3,1)$ and $(3,3)$, as seen in Fig.~\ref{fig.22} for $L_y=\unit[8.1]{nm}$.

%
\begin{figure}
\includegraphics[width=7cm]{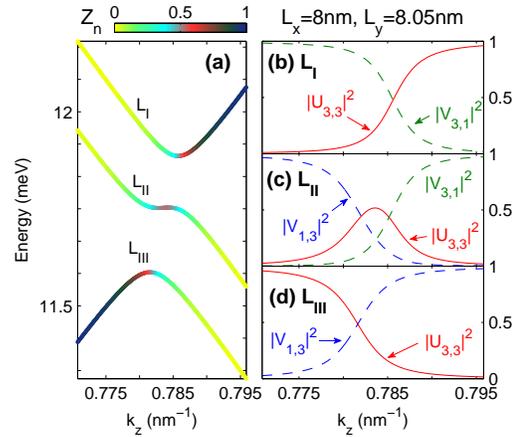}
\caption{(Color online) (a) Band dispersion for the energy bands ($L_I$-$L_{III}$) due to
the Tomasch effect of three-quasiparticles interference of the energy subbands $(1,3)$, $(3,1)$ with
$(3,3)$ for a rectangular cross section $L_x=\unit[8]{nm}$ and $L_y=\unit[8.05]{nm}$. The spectral weight of
the quasiparticle states are indicated by color. The electron and hole components of these states
are shown in panels (b-d), respectively.}\label{fig.23}
\end{figure}
%

An interesting phenomenon is noticed for $L_y=\unit[8.05]{nm}$ where the middle energy band exhibits a flat
region, as seen in Fig.~\ref{fig.22}.  The spectral weight $Z_n$ and the corresponding states [see
Fig.~\ref{fig.23}] show that the Bogoliubov quasiparticle states couple the hole and electron
components.  For the upper energy band $L_I$, the quasiparticle states are converted from the
hole-like states $v_{3,1}$ to the electron-like states $u_{3,3}$ as $k_z$ is increased, as seen in
Fig.~\ref{fig.23}(b).  The same is true for the lower energy band $L_{III}$ but now from the
electron-like states $u_{3,3}$ to the hole-like states $v_{1,3}$, as seen in Fig.~\ref{fig.23}(d).
The middle energy band shows a more complex coupling among the three subbands as it is converted
from $v_{1,3}$ to $v_{3,1}$ with the help of $u_{3,3}$ due to the compatible symmetry of these
states.

\section{Conclusion and discussion}\label{sec:4}

In conclusion, we investigated the Tomasch effect on the electronic structure in nanoscale
superconductors by solving the Bogoliubov-de Gennes equations self-consistently.
Here, the Tomasch effect is induced by an inhomogeneous order parameter appearing due to quantum
confinement.  We found that the Tomasch effect couples degenerate electron and hole states above
the superconducting gap due to quasiparticle interference leading to additional pairs of BCS-like
Bogoliubov-quasiparticles that generate energy gaps resulting in oscillations in the DOS.  When the
energies of the paired states are far from the Fermi level, the pair states show pseudo-gap-like
structures in the DOS.  When they are close to the Fermi level, the pair states result in modulated
wave patterns in the local density of states. All these are due to the inter-subband electron-hole
coupling and their bonding and anti-bonding combinations generating the pair states.

The Tomasch effect is strongly related to the geometrical symmetry of the system and the symmetry,
parity and spacial variation of the order parameter.  For the
nanobelt, the Tomasch effect only leads to two-subband quasiparticles interference processes.  With
even-parity order parameter in the clean limit, the Tomasch effect only plays a role for two states
with the same parity.  A reduced $4\times4$ BdG matrix can describe well the results.  For a
nanowire with a square cross section, the Tomasch effect also results in three-subband quasiparticle
interference processes due to the higher degree of symmetry.  We observe coupling only for the symmetric combination of two hole states, while the anti-symmetric one remains unaffected.
This leads to a unique energy band structure, where one of the subband crosses diagonally across the
induced gap. For nanowires with rectangular cross section, the three-subband quasiparticles interference is
converted into two sets of two-subband quasiparticles interference leading to a distortion of the
previously unaffected band.

The Tomasch effect is commonly formed in inhomogeneous superconductivity but it could be difficult to observe it experimentally.  One reason is that the effect can be shadowed by other states present
at the same energy.  Another reason is that the large size of
Cooper-pairs may result in a complex global electronic structure.   However, the effect can be
enhanced in the following ways: 1) by reducing the symmetry of the sample such as realized by surface
roughness and by making layered junctions; 2) by breaking the symmetry of the order parameter
by e.g. disorder and impurities, and 3) by designing the sample such that the relevant electron
and hole subbands touch each other near their bottom and top, respectively. We have show that for a realistic case, in the presence of weak disorder, the modifications in the DOS due to the TE survive and can be clearly distinguished from oscillations induced by impurity scattering.

\begin{acknowledgments}
This work was supported by the Flemish Science Foundation (FWO-Vlaanderen) and the Methusalem funding of the Flemish Government.
\end{acknowledgments}

\end{document}